\documentclass[a4paper]{article}
\usepackage[english]{babel}
\usepackage{graphicx}

\usepackage[pdfauthor={Giovanni Ramirez Garcia
  (giovanni.ramirez@ecfm.usac.edu.gt)}, pdftitle={Breaking the Area Law: The
  Rainbow State}, pdfsubject={To summarise the results of the rainbow state},
pdfkeywords={entanglement entropy, entanglement spectrum, rainbow
  system}]{hyperref}

\newcommand{\bra}[1]{\left\langle #1 \right|}
\newcommand{\ket}[1]{\left| #1 \right\rangle}

\newcommand{\ketbra}[2]{\left| #1 \right\rangle\!\! \left\langle #2 \right|}

\title{Breaking the Area Law: The Rainbow State}%

\author{Giovanni Ram\'{\i}rez\thanks{Instituto de Investigaci\'on en Ciencias
    F\'{\i}sicas y Matem\'aticas (ICFM-USAC), Universidad de San Carlos de
    Guatemala, Guatemala.  Email: ramirez@ecfm.usac.edu.gt}, Javier
  Rodr\'{\i}guez-Laguna\thanks{Departamento de F\'{\i}sica Fundamental,
    Universidad Nacional de Educacion a Distancia (UNED), Madrid, Spain.},
  Germ\'an Sierra\thanks{Instituto de F\'{\i}sica Te\'{o}rica (UAM-CSIC),
    Madrid, Spain.}}

\date{December, 2018}

\begin{document}
\maketitle

\begin{abstract}
  An exponential deformation of a 1D critical Hamiltonian, with
  couplings falling on a length scale $h^{-1}$, gives rise to ground states
  whose entanglement entropy follows a volume law, i.e. the area law is
  violated maximally. The ground state is now in the so-called {\em rainbow
    phase}, where valence bonds connect sites on the left half with their
  symmetric counterparts on the right.  Here we discuss some of the most
  relevant features of this rainbow phase, focusing on the XX and Heisenberg
  models. Moreover, we show that the rainbow state can be understood either as
  a thermo-field double of a conformal field theory with a temperature
  proportional to $h$ or as a massless Dirac fermion in a curved spacetime
  with constant negative curvature proportional to $h$. Finally, we introduce
  a study of the time-evolution of the rainbow state after a quench to a
  homogeneous Hamiltonian.
\end{abstract}

\section{Introduction}
\label{sec:introduction}

Quantum many-body systems are models which allow us to illustrate important
notions about macroscopic physics, e.g.  magnetic behaviour, in terms of
microscopic elementary interactions between the constituents of that system.
In addition to their physical interest, the development of new methods for
their study has given an impulse to other fields such as quantum integrability
\cite{Baxter1981}, quantum groups \cite{Gomez_etal}, quantum computation and
information \cite{Nielsen_Chuang} or quantum simulators
\cite{Lewenstein_etal2007}.

Early studies of many-body quantum mechanics used to make the assumption that
each particle moves under the effective field created by all the others, i.e.
Hartree-Fock or mean-field type methods \cite{Ashcroft_Mermin}.  These
techniques are very successful to explain many properties of electrons in
solids, through the use of the Fermi liquid approximation or Density
Functional Theory \cite{Hohenberg_Kohn1964, Kohn_Sham1965}.  Nonetheless, they
are unable to take completely into account the effect of strong correlations,
which are a key in many magnetic properties of materials, superconductivity
\cite{BCS1957a, BCS1957b}, quantum Hall effect \cite{Laughlin1981} or
topological insulators \cite{Fu_Kane2007}.

Furthermore, the advent of new technologies such as cold atoms in optical
lattices or trapped ions \cite{Bloch_etal2008, Lewenstein_etal2012}, allows to
engineer quantum systems in which strong correlations are not avoided, but
looked for.  The reasons can be to mimic other quantum systems or to harness
the specific effects of quantum correlations to profit from them, building
better computation and communication technologies.

Quantum entanglement is defined as the property of those pure states which do
not allow a description as product states, i.e. \emph{non-factorizable}
states. For a composite system divided in two parts $A$ and $B$ with Hilbert
space $\mathcal{H} = \mathcal{H}_A \otimes \mathcal{H}_B$, a factorizable
state can be written as $\ket{\psi} = \ket{\psi_A} \otimes \ket{\psi_B}$,
where states $\ket{\psi_A}$ and $\ket{\psi_B}$ describe $A$ and $B$
respectively.  Factorizability of states can be determined using the Schmidt
decomposition \cite{Schmidt1907}, all states in $\mathcal{H}$ can be expressed
as
\begin{displaymath}
  \label{eq:SchmidtDecomp}
  \ket{\psi} = \sum_{i=1}^\chi \lambda_i \ket{a_i} \otimes \ket{b_i} \; ,
\end{displaymath}
where $\ket{a_i}$ and $\ket{b_i}$ are orthonormal states of $\mathcal{H}_A$
and $\mathcal{H}_B$ respectively, and the Schmidt coefficients
$\lambda_i\in \mathbf{R}^+$ and $\sum_i \lambda_i=1$.  The Schmidt number (or
Schmidt rank) $\chi$ is bounded by the dimension of Hilbert spaces of $A$ and
$B$, i.e. $\chi \leq \min \{\dim\{\mathcal{H}_A\}, \dim\{\mathcal{H}_B\}\}$. A
state is factorizable if $\chi=1$, and if $\chi>1$ the state is entangled.

Factorizability defines absence of entanglement. In order to quantify
entanglement it is convenient to consider an observer which is only allowed to
access subsystem $A$. Even when the global state is pure, $\ket{\psi}$, the
subsystem accessible to $A$ can be {\em mixed}. Thus, its quantum-mechanical
description is performed via a reduced density matrix
$\rho_A=\mathrm{Tr}_{\bar A}\ketbra{\psi}{\psi}$, where $\bar A$ is the
complementary of subsystem $A$. We define the von Neumann's entropy of this
reduced density matrix
\begin{displaymath}
  \label{eq:vNEntropy}
  S_A = -\mathrm{tr}\,\{ \rho_A \log{\left(\rho_A\right)} \} \; .
\end{displaymath}

Let us remark that for a pure state, $S_A= S_B$. The von Neumann's entropy
satisfies $S_A\geq 0$, with $S_A= 0$ only for factorizable states. Von
Neumann's entropy is also called the entanglement entropy (EE).

Why is $S_A$ called an entropy? There is a deep relation with the concept of
entropy in statistical mechanics and information theory. Indeed, von Neumann's
entropy is a quantum analogue of Gibbs entropy. But in contrast, it is not
related to thermal fluctuations. The EE can be argued to measure quantum
correlations between subsystems. In classical information theory, the action
of sending a message is also viewed as the action of correlating the sender
and the receiver. The average amount of information contained in that message
is measured with Shannon's entropy \cite{Shannon1948}.  Von Neumann's entropy
is just Shannon's entropy of the eigenvalues of the reduced density matrix,
which can be regarded as a probability distribution. Quantum information
theory builds upon this deep relation between entanglement and information.

The goal of this paper is to present some details of the rainbow model and
then to summarise some results previously obtained.  This work is organised as
follows. First we present the rainbow model and some details of the concentric
valence bond states. After that, we present a summary of the results
previously obtained, we include the references where more details were
discussed.

\section{Concentric Valence Bond States}
\label{sec:cvbs}
Consider a spinless fermion chain of $L$ sites, whose dynamics is described by
the Hamiltonian
\begin{equation}
  \label{eq:FF.H}
  H = - \sum_{i=1}^L t_i \; c^\dagger_i c_{i+1} + \mathrm{h.c.}
\end{equation}
where $t_i$ represents the hopping amplitudes between sites $i$ and $i+1$,
$c_i^\dagger$ and $c_i$ are, respectively, the fermionic creation and
annihilation operators on site $i$, which satisfy anticommutation relations:
$\{ c_j, c^\dagger_k \} = \delta_{jk}$,
$\{ c_j, c_k \} = \{ c_j^\dagger, c^\dagger_k \} = 0$.

Hamiltonian (\ref{eq:FF.H}), which is quadratic in fermionic operators, is
also called free fermion Hamiltonian.  Moreover, free fermion Hamiltonians are
solvable in terms of single-body states that are occupied by particles which
move independently of each other. Diagonalising the hopping matrix,
$T_{ij}= t_i (\delta_{j,i+1}+\delta_{j,i-1})$, $Tv_k= \epsilon_k v_k$, allows
to obtain the single-body energy levels, $\epsilon_k$, and the single-body
modes, $v_{k,i}$, which determine a canonical transformation
\begin{equation}
  \label{eq:FF.Fourier}
  b^\dagger_k =\sum_i v_{k,i} c^\dagger_i \; ,
\end{equation}
where $v_{k,i}$ is an unitary matrix, thus the new operators $b^\dagger_k$
also follow fermionic commutation relations, i.e.  $b_k$ are also fermionic
operators.

All eigenstates of the Hamiltonian (\ref{eq:FF.H}) have the form
\begin{equation}
  \label{eq:FF.GS}
  \ket{\psi}= \prod_{k \in \Omega} b^\dagger_k \ket{0} \; ,
\end{equation}
where $\Omega$ is a subset of $\{1,\cdots,L\}$ denoting the occupied modes and
$\ket{0}$ is the Fock vacuum, which is annihilated by the operators $c_i$. The
energy for the state (\ref{eq:FF.GS}) is $E= \sum_{\Omega}
\epsilon_k$. Therefore, the ground state (GS) is given by filling up all modes
with lowest energy, i.e.  $\Omega = \{ k \mid \epsilon_k<0\}$. In absence of
diagonal terms in the hopping matrix $T$, particle-hole symmetry forces
$\epsilon_k = -\epsilon_{L+1-k}$, so the number of particles of the ground
states is $|\Omega|=n_F=L/2$, i.e. we will operate at half-filling.

The correlation matrix, $C$, has elements defined by
$C_{ij}\equiv \bra{GS} c^\dagger_i c_j \ket{GS}$, which, in terms of the
single-body modes is
\begin{equation}
  \label{eq:CorrMat}
  C_{ij} = \sum_{k=1}^{n_F} \bar v_{k,i} v_{k,j} \; .
\end{equation}

\begin{figure}
  \centering
  \includegraphics[width=.85\textwidth]{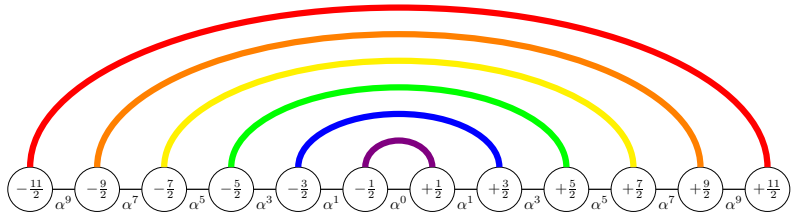}
  \caption[Pictoric representation of the rainbow state]{Rainbow state
    representation, showing the $(-k,+k)$ valence bonds above the
    central link. Each bond contributes as $\log(2)$ to the
    entanglement entropy, thus the entanglement entropy of the left
    (or right) half of the chain is $L \log 2$. The hopping amplitudes
    are given in terms of $\alpha=\exp(-h/2)$.}
  \label{fig:RB.Schemes}
\end{figure}

Let us now describe the family of local Hamiltonians whose GS approaches
asymptotically the rainbow state. For bookkeeping convenience, let us number
the sites as half-integers from $-(L-1)/2$ to $(L-1)/2$, and the corresponding
links as integers, see Fig. \ref{fig:RB.Schemes}. The rainbow Hamiltonian then
reads
\begin{equation}
  \label{eq:RB.H}
  H  \equiv  -\frac{J}{2} c_{\frac{1}{2}}^\dagger c_{-\frac{1}{2}}
  -\frac{J}{2} \sum_{i=\frac{1}{2}}^{L -\frac{3}{2}} e^{-hi} \left[
    c^\dagger_i c_{i+1} +c^\dagger_{-i} c_{-(i+1)} \right] +\mathrm{h.c.}
\end{equation}
where $h\in \mathbf{R}^+$ is the inhomogeneity parameter, and we may also
define $\alpha\equiv \exp(-h/2$, as it is done in
Fig. \ref{fig:RB.Schemes}. Via the Jordan-Wigner transformation, this
Hamiltonian is equivalent to the XX model for a spin-$1/2$ chain. For $h=0$ we
recover the standard uniform 1D Hamiltonian of a spinless fermion model with
open boundary conditions (OBC). Its low energy properties are captured by a
conformal field theory (CFT) with central charge $c = 1$: the massless Dirac
fermion theory, or equivalently (upon bosonization) a Luttinger liquid with
Luttinger parameter $K = 1$.

For $h\gg 1$ we obtain the Hamiltonian used to illustrate a violation of the
area law for local Hamiltonians \cite{Vitagliano_etal2010}.  On the other
hand, for $h<0$ and truncating the chain to the sites $i >0$, one obtains a
Hamiltonian which has the scale-free structure of Wilson's approach to the
Kondo impurity problem \cite{Okunishi_Nishino2010}. Models where $t_i$ is a
hyperbolic function of the site index $i$ were considered in order to enhance
the energy gap \cite{Ueda_etal2010}.

It is worth to notice the striking similarity between our system and the Kondo
chain \cite{Wilson1975}. Indeed, let us divide our inhomogeneous chain into
three parts: central link, left sites and right sites.  The left and right
sites correspond, in our analogy, to the spin up and down chains used in
Wilson's chain representation of the Kondo problem. In both cases, they form a
system of free fermions, with exponentially decaying couplings. In the Kondo
chain, notwithstanding, the central link becomes a magnetic impurity, which
renders the full system non-gaussian.

\section{Results}
\label{sec:results}
In a first paper \cite{Ramirez_etal2014} we analysed the deformation of the
critical local 1D Hamiltonian to explore a smooth crossover between a log law
and a volume law for the EE. We presented the details of the Heisenberg XX
model. The value $h=0$ corresponds to the uniform model, which is described by
the CFT, and in the $h \rightarrow \infty$ limit the GS becomes the rainbow
state.

\begin{figure}
  \centering
  \includegraphics[width=.35\textwidth,angle=-5]{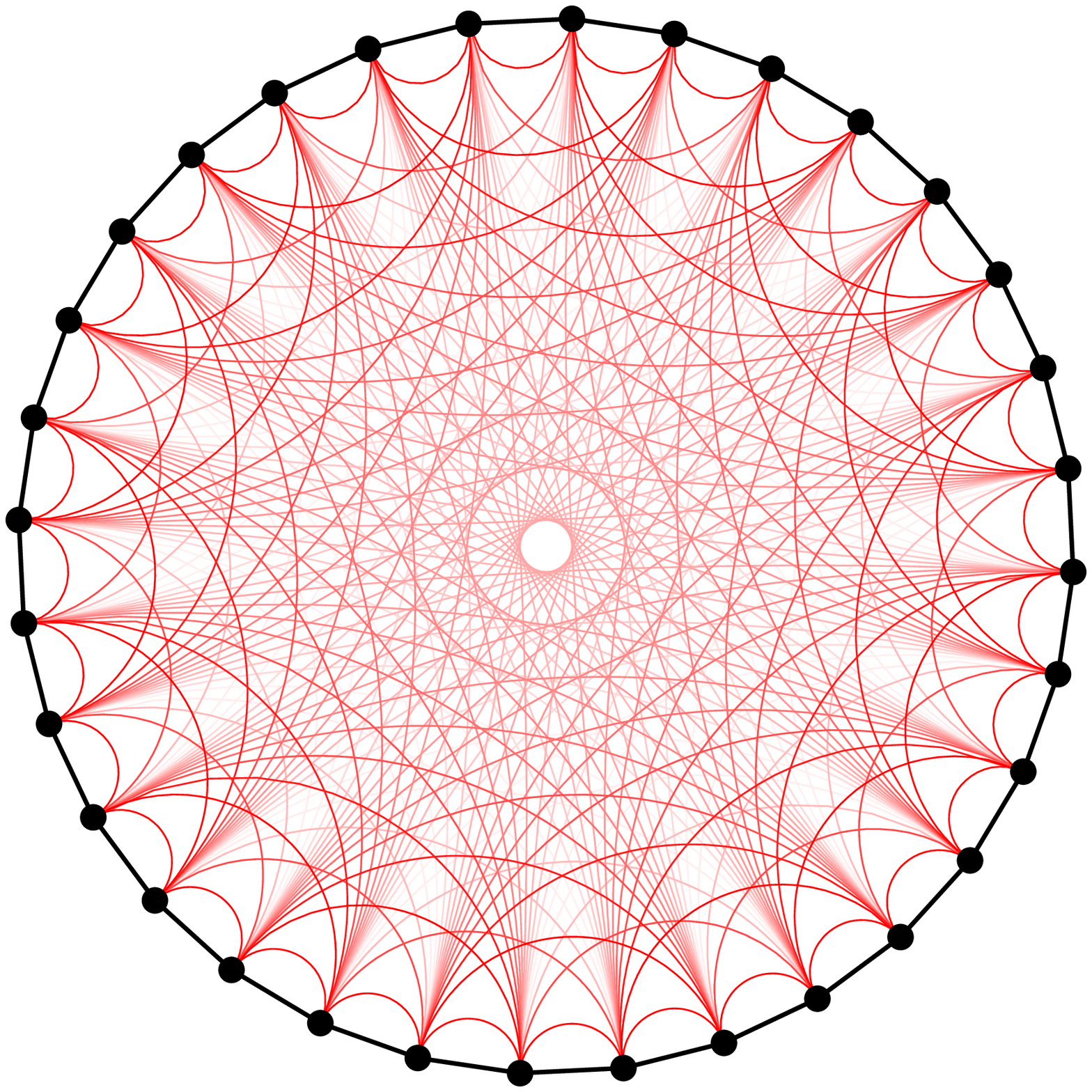}
  \hspace{.02\textwidth}
  \includegraphics[width=.35\textwidth,angle=-5]{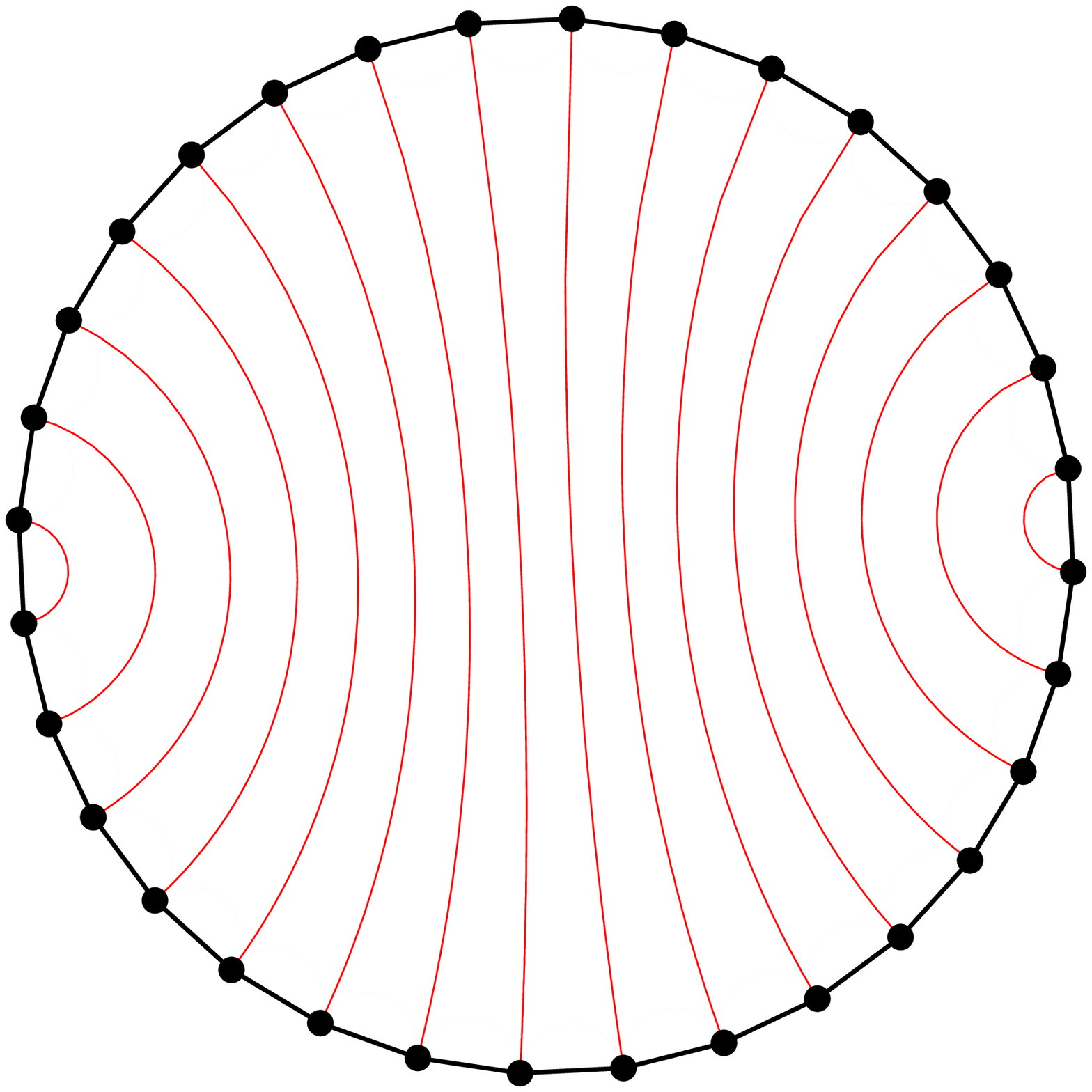}
  \caption[Structure of the correlation matrix]{Structure of the correlation
    matrix defined by the elements $C_{ij}$ for a system of $N=32$ with
    periodic boundary conditions.  \textbf{Left:} for a homogeneous system.
    \textbf{Right:} for a rainbow system.}
  \label{fig:Corr.Structs}
\end{figure}

We used a graphical representation for the correlation matrix which operates
as follows. For any matrix element, $C_{ij}$, we draw a line inside the unit
circle with a colour which marks the strength of the correlation between those
points. A finitely-correlated state will be characterised by a correlation
matrix whose representation is given by short lines which do not go deep
inside the unit circle. A conformal state, with infinite correlation length,
is characterised by a certain self-similar structure in the geodesic
pattern. Realisations of the rainbow state correlations are also very easy to
spot. Figure \ref{fig:Corr.Structs} shows the structure of the correlation
matrix for a 1D system with periodic boundary conditions (PBC) for a different
systems. The intensity of each line connecting two sites is related to
$\left| C_{ij} \right|$.

\begin{figure}
  \centering
  \includegraphics[width=.75\textwidth]{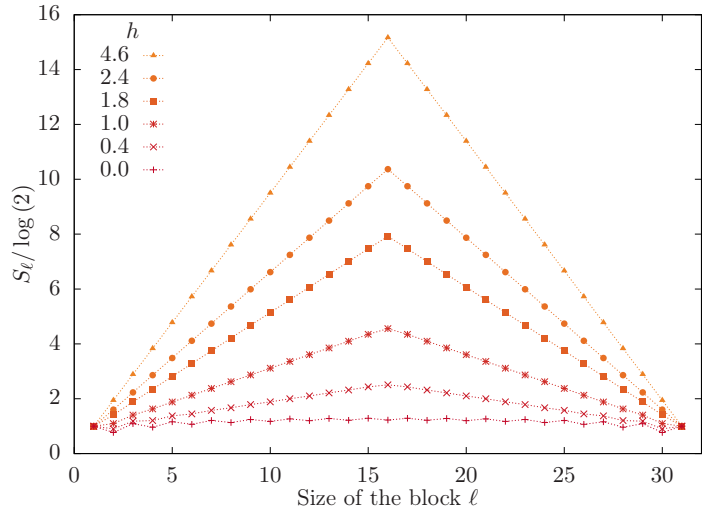}
  \caption[Block entropy for the rainbow state]{Block entropy $S_\ell$, for a
    system of size $L=16$ ($32$ sites).  Notice the \emph{tent shape} appear
    for small inhomogeneities in the system, denoting volumetric growth of the
    entanglement entropy.}
  \label{fig:RB.blockEntropy}
\end{figure}

We studied numerically the EE of blocks containing $\ell$ sites starting from
the left extreme of the chain within the GS of Hamiltonian
(\ref{eq:RB.H}). For large values of $h$ we observe a characteristic
\emph{tent shape} in the EE, i.e. an approximately linear growth up to
$\ell=L$ followed by a symmetric linear decrease, giving the volumetric
behaviour. Figure \ref{fig:RB.blockEntropy} shows the EE for different values
of $h$ for a system of $N=32$ sites.  As the value of $h$ decreases, the slope
decreases and ripples start to appear and for $h=0$ they recover the parity
oscillations characteristics of the von Neumann entropy with OBC. For positive
$h$, the slope was empirically shown to be given by $h/6$.

The analysis of the entanglement spectrum showed very interesting connections
between conformal growth $S \sim \log(L)$ and volumetric growth $S \sim
L$. Indeed, the spectrum is approximately equally spaced, with an entanglement
spacing $\Delta L$ that decays with the system size as $1/\log(L)$ at the
conformal point and as $1/L$ for rainbow system. We have also found that the
EE is approximately proportional to the inverse of the entanglement spacing,
in wide regions of the parameter space, which generalises previous known
results for critical and massive systems.

In a second paper we have extended the analysis using field-theory methods
\cite{Ramirez_etal2015}. We showed that the system can be described as a
conformal deformation of the homogeneous case $h=0$, given by the following
transformation
\begin{equation}
  \label{eq:equis}
  \tilde x = \mathrm{sign}(x) \frac{e^{h|x|}-1}{h} \; ,
\end{equation}
which maps the interval $x\in[-L,L]$ to $\tilde x \in [-\tilde L, \tilde L]$,
where $\tilde L = (e^{hL}-1)/h$.

\begin{figure}
  \centering
  \includegraphics[width=.75\textwidth]{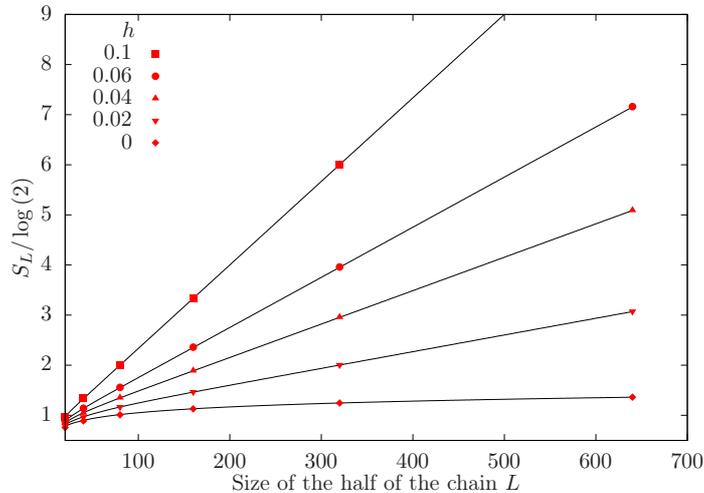}
  \caption[Half-chain entropy obtained by the continuum
  approximation]{Comparing numerical results of the half-chain entropy of the
    rainbow system for different values of $h$ with the theoretical
    prediction.}
  \label{fig:S_prediction}
\end{figure}

If we expand the local operators $c_n$ into slow modes, $\psi_R(x)$ and
$\psi_L(x)$ around the Fermi points $\pm k_F$
\begin{equation}
  \label{eq:AnihilationExpansion}
  \frac{c_n}{\sqrt{a}} \simeq e^{ik_F x} \psi_L(x) + e^{-ik_F x}\psi_R(x) \; ,
\end{equation}
located at the position $x=an \in (-\mathcal{L}, \mathcal{L})$, where $a$ is
the lattice spacing and $\mathcal{L}=aL$ and, at half-filling $k_F=\pi/(2a)$
is the Fermi momentum. With this expansion we were able to obtain the
Hamiltonian
\begin{equation}
  \label{eq:TransformedHamiltonian}
  H \simeq i \int_{-\tilde L}^{\tilde L} d \tilde x \left[ \tilde
    \psi_R^\dagger \partial_{\tilde x} \tilde\psi_R - \tilde\psi_L^\dagger
    \partial_{\tilde x} \tilde \psi_L\right] \; ,
\end{equation}
which represents the free fermion Hamiltonian for a chain of length
$2\tilde L$ under transformation (\ref{eq:equis}) with the the fermion fields
given by
\begin{equation}
  \label{eq:fermionFields}
  \tilde \psi_{R,L} (\tilde x) = \left( \frac{d \tilde x}{d x}
  \right)^{-1/2} \psi_{R,L} (x) \; .
\end{equation}
Thus, using conformal invariance and substituting $L$ by $\tilde L$, we were
able to transform the EE for a block of one-half of the critical system
\cite{Calabrese_Cardy2004}
\begin{equation}
  \label{eq:Scft}
  S^{CFT}_L=c \log(L)/6 +c' \; ,
\end{equation}
into the EE of the rainbow system
\begin{equation}
  \label{eq:rainbowEntropy}
  S_L =\frac{c}{6} \log{\left( \frac{e^{hL}-1}{h} \right)} + c' \; ,
\end{equation}
which is shown in Figure \ref{fig:S_prediction} as a function of the
half-chain size to check the theoretical prediction for different values of
$h$.

We also showed that the corresponding conformal transformations suggests the
definition of a temperature $T=1/\beta=h/(2\pi)$.  Based on the entanglement
spectrum, which is given the eigenvalues of the entanglement Hamiltonian
\begin{equation}
  \label{eq:entHam}
  H_E = \sum_p \epsilon_p b^\dagger_pb_p + f_0 \; ,
\end{equation}
where $b_p$ and $b^\dagger_p$ are fermion operators and the entanglement
energies are given approximately by
\begin{equation}
  \label{eq:EEnergies}
  \epsilon_p \simeq \Delta_L p, \quad \mathrm{for}\;   p = \left\{
  \begin{array}{ll} 
    \pm \frac{1}{2}, \pm \frac{3}{2}, \ldots, \pm \frac{L-1}{2}, & L:
                                                                  \mathrm{even}\;
                                                                  ,\\ 
    0, \pm 1, \pm 2, \ldots, \pm (L-2), & L: \mathrm{odd} \; ,
  \end{array}\right.
\end{equation}
where the level spacing $\Delta_L$ is related to the half-chain entropy.
Moreover, the single-body entanglement energies fulfil
\begin{equation}
  \label{eq:relation}
  \epsilon_p \simeq \beta \epsilon^{CFT}_p = \left( \frac{2\pi}{h} \right)
  \left( \frac{\pi}{L} p\right) = \frac{2\pi^2}{z} p \; .
\end{equation}
Thus, the appearance of a volume law entropy is linked to the existence of an
effective temperature for the GS of the rainbow model that was finally
identified with a thermo-field state
\begin{equation}
  \label{eq:thermo}
  \ket{\psi} = \sum_n e^{-\beta E_n/2} \ket{n}_L \ket{n}_R \; ,
\end{equation}
where $\ket{n}_R$ and $\ket{n}_L$ correspond to the homogeneous GS for the
right and left parts of the chain with $h=0$. This striking result points
towards an unexpected connection with the theory of black holes and the
emergence of space-time from entanglement. These intriguing connections were
further explored within the framework of CFT \cite{Tonni_2018}. Furthermore,
we also showed how the deformation on the system accounts for the change in
the dispersion relation, the single-particle wave functions in the vicinity of
the Fermi point and the half-chain von Neumann and R\'enyi entropies. Finally,
we show how to extend the rainbow Hamiltonian to more dimensions in a natural
way and we to checked that the EE of the 2D analogue grows as the area of the
block.

In a third paper \cite{Rodriguez_etal2017} we applied methods of 2D quantum
field theory in curved space-time to determine the entanglement structure of
the rainbow phase. We showed that the rainbow system can be described by a
massless Dirac fermion on a Riemannian manifold with constant negative
curvature everywhere except at the centre, equivalent to a Poincar\'e metric
with a strip removed.  We used this identification to apply CFT for
inhomogeneous 1D quantum systems \cite{Dubail_etal2017} to provide accurate
predictions for the smooth part of the $n$-order R\'enyi EE for blocks of
different types such as the block at the chain's edge given by the bipartition
$A=[-L,x]$, $B=[x,L]$
\begin{equation}
  \label{eq:EEleft}
  S^{CFT}_n(x) = \frac{n+1}{12n} \ln{Y(x)} \; ,
\end{equation}
where
\begin{equation}
  \label{eq:Yleft}
  Y(x) = 8 e^{-h|x|} \frac{e^{hL} -1}{\pi L} \cos{\left( \frac{\pi}{2}
      \frac{e^{h|x|} -1}{e^{hL} -1} \right)} \; ,
\end{equation}
or for a block at an arbitrary position given by the bipartition
$A=[x_1,x_2]$, $B=[-L,x_1] \cup [x_2,L]$
\begin{equation}
  \label{eq:EEIIIarb}
  S^{CFT}_n (x_1,x_2) = \frac{n+1}{12n} \ln{4Y(x_1,x_2)} + E_n \; ,
\end{equation}
where $E_n$ is a non-universal constant, and with $e^\sigma=e^{-h|x|}$ we have
\begin{equation}
  \label{eq:Yarb}
  Y(x_1,x_2) = \frac{e^{\sigma(x_1)+\sigma(x_2)} \, 16\tilde L^2}{\pi^2
    \cos{\left( \frac{\pi (\tilde x_1 +\tilde x_2)}{4\tilde L}\right)}}
  \sin{\left( \frac{\pi (\tilde x_1 - \tilde x_2)}{4\tilde L} \right)^2}
  \cos{\left( \frac{\pi\tilde x_1}{2\tilde L} \right)} \cos{\left( \frac{\pi
        \tilde x_2}{2\tilde L} \right)}\; .
\end{equation}

\begin{figure}
  \centering
  \includegraphics[width=.75\textwidth]{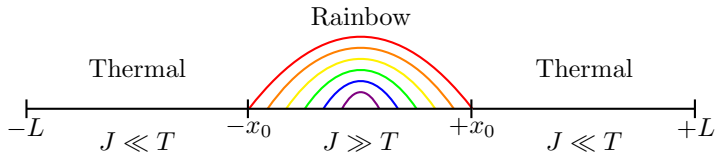}
  \caption[Half-chain entropy obtained by the continuum
  approximation]{Illustration for the finite temperature behaviour of the
    rainbow chain. Let us find $x_0$ such that $T\approx J(x_0)
    =J(-x_0)$. Then, for $|x| > |x_0|$, $J(x)\ll T$ and the system behaves as
    if it were at infinite temperature. On the other hand, for $|x| < |x_0|$
    the system behaves as if at zero temperature.}
  \label{fig:finiteT}
\end{figure}

Furthermore, we also showed that for a physical temperature $T=J(x_0)$, i.e. a
temperature in the range of the energies spanned by the values of the hopping
amplitude for a point $x_0 \in [-L,L]$, the system splits into three regions:
$x<-x_0$, $x\in [-x_0,x_0]$ and $x>x_0$ (cf. Figure \ref{fig:finiteT}).  The
central region still behaves as if it were at $T=0$ while the two extremes as
if they were at $T\rightarrow \infty$.  Thus, the entropy of a block is
obtained by adding the contributions on each region
\begin{equation}
  \label{eq:fullEE}
  S(x) \sim \left\{
    \begin{array}{ll}
      (L-|x|)\ln 2, & x \in (-L,-x_0) \; , \\
      (L-x_0)\ln 2 + (x_0-|x|) h/6, & x \in (-x_0,x_0) \; , \\
      (L- 2x_0 +x) \ln 2, & x \in (x_0,L) \;.
    \end{array}\right.
\end{equation}

\begin{figure}
  \centering
  \includegraphics[width=.45\textwidth]{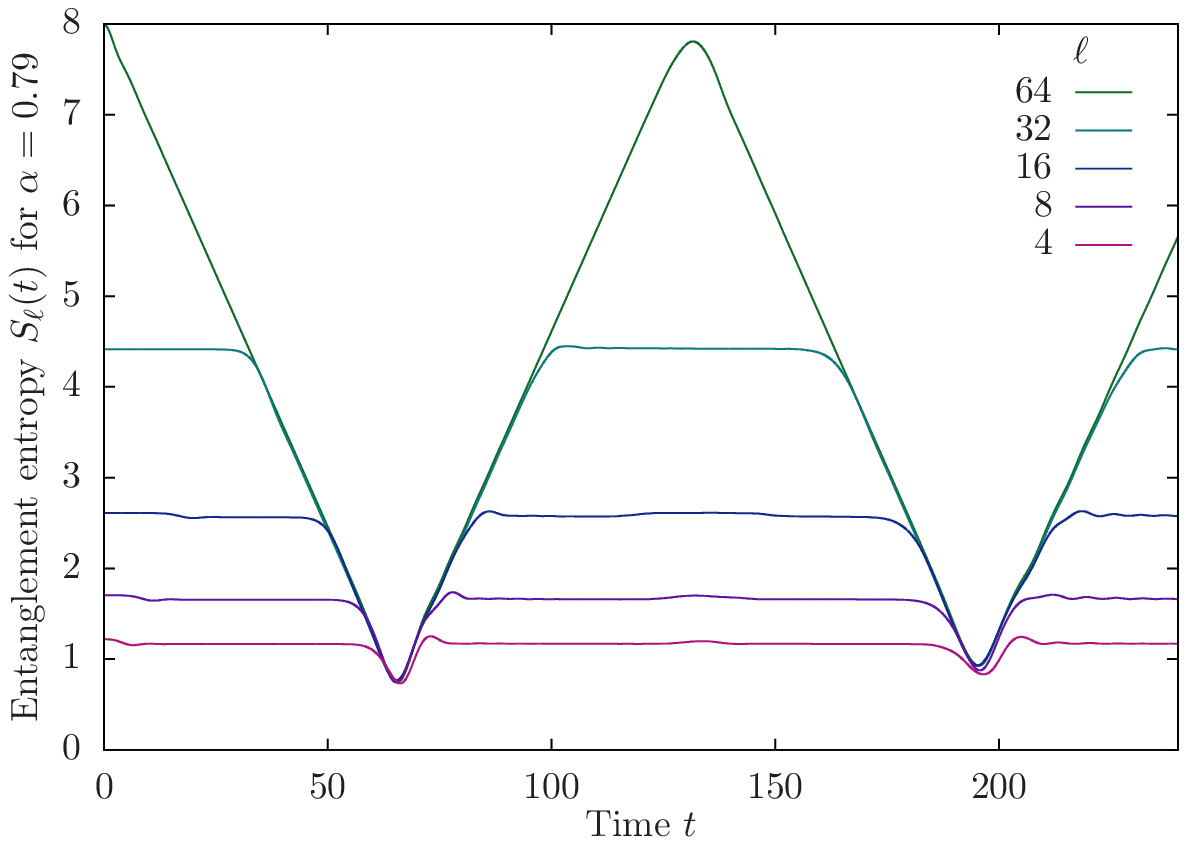}
  \hspace{.01\textwidth}%
  \includegraphics[width=.45\textwidth]{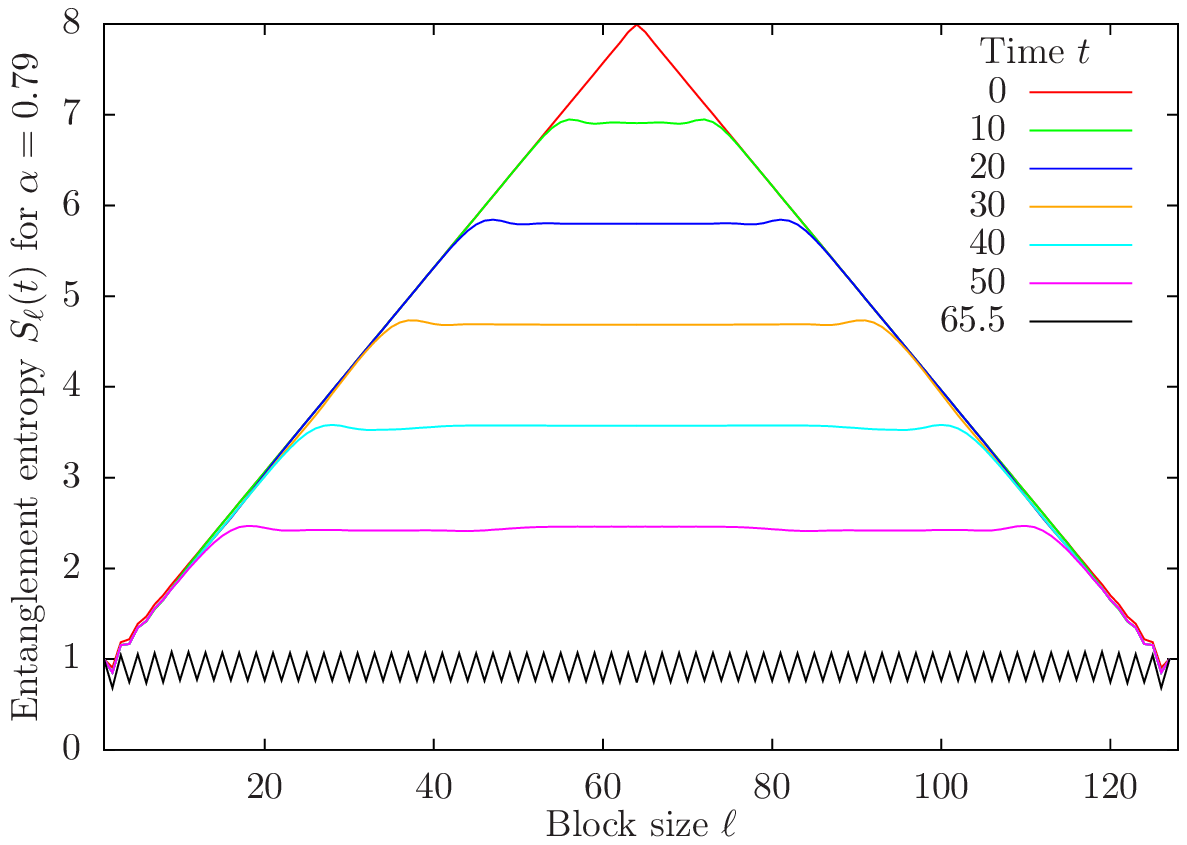}
  \caption[Time evolution of leftmost blocks for the rainbow state]{Time
    evolution of the entanglement entropy.  \textbf{Left:} for blocks of size
    $\ell$ starting from the leftmost site. Notice the time periodicity and
    the strict order: the entropy of a larger block is always strictly larger
    than the entropy of a smaller one. \textbf{Right:} for the tent shape
    (cf. Fig. \ref{fig:RB.blockEntropy}) of the entropy.}
  \label{fig:LeftBlocks}
\end{figure}

In a fourth paper \cite{ramirez_etal2019}, we study the time-evolution of the
rainbow state and the dimer state, after a quench to a homogeneous Hamiltonian
in 1D. The subsequent evolution of the EE presents very intriguing
features. First, the EE of the half-chain of the rainbow decreases linearly
with time and, after it reaches a minimal value, it increases again,
eventually reaching (approximately) the initial state.  Blocks of smaller
sizes only decrease after a certain transient time, which can not be explained
directly via a Lieb-Robinson bound \cite{Lieb_Robinson1971}, since the quench
is global (cf. Figure \ref{fig:LeftBlocks}).

The dimer state, on the other hand presents an opposite behaviour: the EE
grows linearly for all blocks, reaching a maximally entangled state which
resembles the rainbow state. Afterwards, the EE decreases again, cyclically.
We also focus on the correlation between pairs of sites which suggests the
motion of quasiparticles. Thus, we will attempt a theoretical explanation in
terms of an extension of the quasiparticle picture by Calabrese and Cardy
\cite{Calabrese_Cardy2005}.

\textbf{Acknowledgements}.  We would like to thank Pasquale Sodano for
organising the conferences hosted at the International Institute of Physics
(Natal) and for his efforts to collect the contributions for the Proceedings.

\bibliographystyle{unsrt}
\bibliography{references}

\begin{thebibliography}{10}

\bibitem{Baxter1981}
R.~J. Baxter.
\newblock {Corner transfer matrices}.
\newblock {\em Physica A.}, 106(1--2):18--27, 1981.

\bibitem{Gomez_etal}
C.~G\'omez, M.~Ruiz-Altaba, and G.~Sierra.
\newblock {\em {Quantum Groups in Two-Dimensional Physics}}.
\newblock Cambridge Lecture Notes in Physics. Cambridge University Press, 1996.

\bibitem{Nielsen_Chuang}
M.~A. Nielsen and I.~L. Chuang.
\newblock {\em {Quantum Computation and Quantum Information: 10th Anniversary
  Edition}}.
\newblock Cambridge University Press, 10 edition, 2010.

\bibitem{Lewenstein_etal2007}
M.~Lewenstein, A.~Sanpera, V.~Ahufinger, B.~Damski, A.~Sen(De), and U.~Sen.
\newblock {Ultracold atomic gases in optical lattices: mimicking condensed
  matter physics and beyond}.
\newblock {\em Adv. Phys.}, 56(2):243--379, 2007.

\bibitem{Ashcroft_Mermin}
N.~W. Ashcroft and N.~D. Mermin.
\newblock {\em {Solid State Physics}}.
\newblock Harcourt College Publishers, 1976.

\bibitem{Hohenberg_Kohn1964}
P.~Hohenberg and W.~Kohn.
\newblock {Inhomogeneous Electron Gas}.
\newblock {\em Phys. Rev.}, 136:B864--B871, Nov 1964.

\bibitem{Kohn_Sham1965}
W.~Kohn and L.~J. Sham.
\newblock {Self-Consistent Equations Including Exchange and Correlation
  Effects}.
\newblock {\em Phys. Rev.}, 140:A1133--A1138, Nov 1965.

\bibitem{BCS1957a}
J.~Bardeen, L.~N. Cooper, and J.~R. Schrieffer.
\newblock {Microscopic Theory of Superconductivity}.
\newblock {\em Phys. Rev.}, 106:162--164, Apr 1957.

\bibitem{BCS1957b}
J.~Bardeen, L.~N. Cooper, and J.~R. Schrieffer.
\newblock {Theory of Superconductivity}.
\newblock {\em Phys. Rev.}, 108:1175--1204, Dec 1957.

\bibitem{Laughlin1981}
R.~B. Laughlin.
\newblock {Quantized Hall conductivity in two dimensions}.
\newblock {\em Phys. Rev. B}, 23:5632--5633, May 1981.

\bibitem{Fu_Kane2007}
L.~Fu and C.~L. Kane.
\newblock {Topological insulators with inversion symmetry}.
\newblock {\em Phys. Rev. B}, 76:045302, Jul 2007.

\bibitem{Bloch_etal2008}
I.~Bloch, J.~Dalibard, and W.~Zwerger.
\newblock {Many-body physics with ultracold gases}.
\newblock {\em Rev. Mod. Phys.}, 80:885--964, Jul 2008.

\bibitem{Lewenstein_etal2012}
M.~Lewenstein, A.~Sanpera, and V.~Ahufinger.
\newblock {\em {Ultracold Atoms in Optical Lattices: Simulating quantum
  many-body systems}}.
\newblock Oxford University Press, 2012.

\bibitem{Schmidt1907}
E.~Schmidt.
\newblock {Zur Theorie der linearen und nichtlinearen Integralgleichungen}.
\newblock {\em Math. Ann.}, 63(4):433--476, 1907.

\bibitem{Shannon1948}
C.E. Shannon.
\newblock {A mathematical theory of communication}.
\newblock {\em The Bell System Technical Journal}, 27(3):379--423, July 1948.

\bibitem{Vitagliano_etal2010}
G.~Vitagliano, A.~Riera, and J.~I. Latorre.
\newblock {Volume-law scaling for the entanglement entropy in spin-1/2 chains}.
\newblock {\em New J. Phys.}, 12(11):113049, 2010.

\bibitem{Okunishi_Nishino2010}
K.~Okunishi and T.~Nishino.
\newblock {Scale-free property and edge state of Wilson's numerical
  renormalization group}.
\newblock {\em Phys. Rev. B}, 82:144409, Oct 2010.

\bibitem{Ueda_etal2010}
H.~Ueda, H.~Nakano, K.~Kusakabe, and T.~Nishino.
\newblock {Scaling Relation for Excitation Energy under Hyperbolic
  Deformation}.
\newblock {\em Progr. Theor. Phys.}, 124(3):389--398, 2010.

\bibitem{Wilson1975}
K.~G. Wilson.
\newblock {The renormalization group: Critical phenomena and the Kondo
  problem}.
\newblock {\em Rev. Mod. Phys.}, 47:773--840, Oct 1975.

\bibitem{Ramirez_etal2014}
G.~Ram\'{\i}rez, J.~Rodr\'{\i}guez-Laguna, and G.~Sierra.
\newblock {From conformal to volume law for the entanglement entropy in
  exponentially deformed critical spin 1/2 chains}.
\newblock {\em J. Stat. Mech.}, 2014(10):P10004, 2014.

\bibitem{Ramirez_etal2015}
G.~Ram\'{\i}rez, J.~Rodr\'{\i}guez-Laguna, and G.~Sierra.
\newblock {Entanglement over the rainbow}.
\newblock {\em J. Stat. Mech.}, 2015(6):P06002, 2015.

\bibitem{Calabrese_Cardy2004}
P.~Calabrese and J.~Cardy.
\newblock Entanglement entropy and quantum field theory.
\newblock {\em J. Stat. Mech.}, 2004:P06002, 2004.

\bibitem{Tonni_2018}
E.~Tonni, J.~Rodriguez-Laguna, and G.~Sierra.
\newblock {Entanglement hamiltonian and entanglement contour in inhomogeneous
  1D critical systems}.
\newblock {\em J. Stat. Mech.}, 2018:043105, 2018.

\bibitem{Rodriguez_etal2017}
J.~Rodr{\'{\i}}guez-Laguna, J.~{Dubail}, G.~Ram{\'{\i}}rez, P.~Calabrese, and
  G.~Sierra.
\newblock {More on the rainbow chain: entanglement, space-time geometry and
  thermal states}.
\newblock {\em J. Stat. Mech.}, 50(16):164001, 2017.

\bibitem{Dubail_etal2017}
J.~Dubail, J.-M. St\'ephan, J.~Viti, and P.~Calabrese.
\newblock {Conformal Field Theory for Inhomogeneous One-dimensional Quantum
  Systems: the Example of Non-Interacting Fermi Gases}.
\newblock {\em SciPost Phys.}, 2:002, 2017.

\bibitem{ramirez_etal2019}
G.~Ram\'{\i}rez, J.~Rodr\'{\i}guez-Laguna, and G.~Sierra.
\newblock Quenched dynamics of valence bond states.
\newblock Work in progress.

\bibitem{Lieb_Robinson1971}
E.~H. Lieb and D.~W. Robinson.
\newblock {The finite group velocity of quantum spin systems}.
\newblock {\em Commun. Math. Phys.}, 28(3):251--257, 1972.

\bibitem{Calabrese_Cardy2005}
P.~Calabrese and J.~Cardy.
\newblock {Evolution of entanglement entropy in one-dimensional systems}.
\newblock {\em J. Stat. Mech.}, 2005(04):P04010, Apr 2005.

\end{thebibliography}

\end{document}